\documentclass[showpacs,aps,pre,preprint]{revtex4}

\usepackage{amsmath}
\usepackage{graphicx}
\usepackage{dcolumn}
\usepackage{bm}

\begin{document}

\title{Shear state of freely evolving granular gases}
\author{J. Javier Brey, M.J. Ruiz-Montero, and A. Dom\'{\i}nguez}
\affiliation{F\'{\i}sica Te\'{o}rica, Universidad de Sevilla,
Apartado de Correos 1065, E-41080, Sevilla, Spain}
\date{\today }

\begin{abstract}
Hydrodynamic equations are used to identify the final state reached
by a freely evolving granular gas above but close to its shear
instability. The theory predicts the formation of a two bands shear
state with a steady density profile. There is a modulation between
temperature and density profiles as a consequence of the energy
balance, the density fluctuations remaining small, without producing
clustering. Moreover, the time dependence of the velocity field can
be scaled out with the squared root of the average temperature of
the system. The latter follows the Haff law, but with an effective
cooling rate that is smaller than that  of the free homogeneous
state. The theoretical predictions are compared with numerical
results for inelastic hard disks obtained by using the direct Monte
Carlo simulation method, and a good agreement is obtained for low
inelasticity.

\end{abstract}

\pacs{45.70.Mg, 45.70.Qj, 47.20.Ky} \maketitle

\section{Introduction}
\label{s1} Granular gases are ensembles of macroscopic particles
whose dynamics is controlled by inelastic binary collisions,
separated by ballistic motion \cite{Ca90}. The forces are usually
short range and repulsive and, therefore, in the simplest form
granular gases are modeled as a collection of smooth inelastic hard
spheres or disks. In recent years, the study of granular gases has
attracted a lot of attention \cite{ByP04,Du00}, among other reasons
because they serve as a sensitive proving ground for kinetic theory
and non-equilibrium statistical mechanics.

A peculiarity of granular gases is their tendency to spontaneously
develop pattern forming instabilities when evolving freely. This
includes the so-called shearing and clustering  instabilities
\cite{GyZ93,McyY94}, in which the system develops patterns both in
the flow field and the density field. Analysis of the hydrodynamic
Navier-Stokes equations for granular gases indicates the existence
of several possible scenarios explaining the development of these
instabilities and their mutual influence. An illuminating discussion
of them is given in ref. \cite{PAFyM08}. For instance, the existence
of non-stationary channel flows in freely cooling gases leading to
an attempted finite-time density blowup, which is arrested by heat
diffusion, has been shown recently \cite{ELyM05,MFyV08}.

In this paper, a shear flow state of freely evolving dilute granular
gases will be described. Its origin is tied in with the shear mode
instability and the analysis to be presented here is restricted to
systems whose size is slightly larger than the critical size for
this instability. The latter happens to be smaller than the critical
size for the instability of one of the longitudinal modes associated
with the density \cite{BDKyS98,WByE02}. Consequently, the spatial
inhomogeneities of the hydrodynamic fields are expected to be small.
Although the flow is non-stationary, all the time dependence can be
expressed through the spatially averaged temperature and, therefore,
it can be described in terms of dimensionless time-independent
quantities. More precisely, it is an inhomogeneous two band shear
state. It must be stressed that this state is exhibited by a system
with, for instance, periodic or elastic boundary conditions, without
being driven by the boundaries as it happens with the so-called
simple or uniform shear flow \cite{Lu84,JyR88,SGyN96}.
Non-stationary states in which all the time dependence occurs
through the temperature, like the homogenous cooling state
\cite{Ha83}, as well as non-uniform states where all the spatial
dependence also occurs through the temperature are peculiar of
granular gases \cite{BCMyR01}.

The existence of this free shear state has been suggested previously
in the framework of a time-dependent Ginzburg-Landau model for
granular gases \cite{WByE02}, and also as the result of a nonlinear
analysis of the shearing instability \cite{SMyM00}. The theory
presented in this paper differs from those studies both in the
method and the results, where rather strong discrepancies occur as
it will be discussed along the paper. The interest here is focussed
on the complete identification of the hydrodynamic fields
characterizing the free shear state and the comparison of the
theoretical predictions with the results of numerical simulations of
the particles composing the system. Of course, this provides a very
demanding test of the validity of the macroscopic description
provided by the hydrodynamic Navier-Stokes equations for granular
gases.

The remainder of the paper is organized as follows. In Sec.\
\ref{s2}, the nonlinear Navier-Stokes hydrodynamic equations for a
dilute granular gas are shortly reviewed, and particularized for the
free shear state. This is macroscopically characterized by
presenting gradients only in one direction and by a velocity field
perpendicular to the gradients. Moreover, it is assumed that the
density inhomogeneities are small. By introducing appropriate
dimensionless length and time scales, a closed equation for the
velocity flow is derived in Sec.\ \ref{s3}. The solution of this
equation shows that the amplitude of the  velocity field decays
monotonically in time. Moreover, by analyzing the temperature
equation, an explicit expression for the amplitude of the steady
density fluctuation is obtained. Also, it is seen that the ratio
between the average temperature of the system and the square of the
amplitude of the velocity field is time independent.

In Sec. \ref{s4}, the simulation method used is indicated. It is
restricted to low density gases, whose one-particle distribution
function obeys the Boltzamnn equation, but it must be stressed that
it is an $N$-particle simulation method, therefore providing
information on all the space and time correlations. It is observed
that the system actually tends to a steady situation having the
assumed properties of the free shear state. Then, the measured
hydrodynamic quantities are compared with the analytical theoretical
expressions derived from the Navier-Stokes equations. A good
agreement is obtained for small inelasticity and system sizes close
to the critical value for the shear instability. Finally, Sec.
\ref{s5} summarizes the obtained results and presents a brief
discussion.

\section{Hydrodynamic equations}
\label{s2}

The balance equations for the number density $n({\bm r},t)$, the
flow velocity ${\bm u}({\bm r},t)$, and the granular temperature
$T({\bm r},t)$ of a system composed by smooth inelastic hard spheres
($d=3$) or disks ($d=2$) of mass $m$, diameter $\sigma$, and
coefficient of normal restitution $\alpha$, have the form
\begin{equation}
\frac{ \partial n}{\partial t}+{\bm \nabla}\cdot \left( n {\bm
u}\right) =0, \label{2.1}
\end{equation}
\begin{equation}
\left( \frac{\partial}{\partial t}+ {\bm u} \cdot {\bm
\nabla}\right) {\bm u}+(mn)^{-1} {\bm \nabla} \cdot \mathsf{P} =0,
\label{2.2}
\end{equation}
\begin{equation}
\left( \frac{\partial}{\partial t}+ {\bm u} \cdot {\bm \nabla} +
\zeta \right) T + \frac{2}{nd}\left( {\sf P} : {\bm \nabla} {\bm
u}+{\bm \nabla}\cdot {\bm q}\right) =0. \label{2.3}
\end{equation}
For a dilute gas and to Navier-Stokes order, the pressure tensor
$\mathsf{P}$, the heat flux ${\bm q}$, and the cooling rate $\zeta$
are given by the constitutive relations \cite{BDKyS98,ByC01,Go03}
\begin{equation}
\mathsf{P}_{ij}({\bm r},t)= p({\bm r},t) \delta_{ij}-\eta \left(
\frac{\partial u_{j}}{\partial r_{i}} +\frac{\partial
u_{i}}{\partial r_{j}} -\frac{2}{d}\delta _{ij} {\bm \nabla}\cdot
{\bm u}\right), \label{2.4}
\end{equation}
\begin{equation}
\mathbf{q}(\mathbf{r},t) = -\kappa {\bm \nabla}T-\mu {\bm \nabla}n,
\label{2.5}
\end{equation}
\begin{equation}
\zeta=\zeta^{(0)}+ \zeta^{(2)}. \label{2.6}
\end{equation}
Here $p=nT$ is the hydrostatic pressure, $\eta$ is the shear
viscosity, $\kappa$ the (thermal) heat conductivity, and $\mu$ a
transport coefficient peculiar of granular fluids that is referred
to as the diffusive heat conductivity. The term $\zeta^{(0)}$
denotes the contribution to the cooling rate of zeroth order in the
gradients of the hydrodynamic fields, while $\zeta^{(2)}$ denotes
the second order in the gradients contributions. The latter will be
neglected in the following, as it is done in most of the
calculations, since it is expected to give very small corrections to
the hydrodynamic equations as compared with the similar terms coming
from the pressure tensor and the heat flux \cite{BDKyS98}. The
transport coefficients appearing in Eqs. (\ref{2.4}) and (\ref{2.5})
can be expressed as
\begin{equation}
\label{2.7} \eta(T,\alpha) = \eta_{0}(T) \eta^{*}(\alpha),
\end{equation}
\begin{equation}
\label{2.8} \kappa(T,\alpha)=\kappa_{0}(T) \kappa^{*}(\alpha),
\end{equation}
\begin{equation}
\label{2.9} \mu (n,T, \alpha)= \frac{T \kappa_{0}(T)}{n}\,
\mu^{*}(\alpha),
\end{equation}
where $\eta_{0}(T)$ and $\kappa_{0}(T)$ are the low density
(Boltzmann) values of the shear viscosity and the thermal heat
conductivity, respectively. Their explicit expressions as well as
those of the dimensionless functions $\eta^{*}$, $\kappa^{*}$, and
$\mu^{*}$ are given in Appendix \ref{ap1}. Finally, the expression
of the zeroth order cooling rate reads
\begin{equation}
\label{2.10} \zeta^{(0)}(n,T,\alpha)= \frac{n T}{\eta_{0}(T)}\,
\zeta^{*}(\alpha).
\end{equation}
The function $\zeta^{*}(\alpha)$ is also given in Appendix
\ref{ap1}.

The hydrodynamic Navier-Stokes equations for the inelastic gas  are
obtained by substituting Eqs. (\ref{2.4})-(\ref{2.6}) into Eqs.
(\ref{2.1})-(\ref{2.3}). They admit a simple solution defined by a
constant and uniform density $n_{H}$, a vanishing velocity field
${\bm u}_{H}={\bm 0}$, and a time dependent temperature $T_{H}(t)$
obeying the equation
\begin{equation}
\label{2.11} \frac{\partial T_{H}(t)}{\partial t}+
\zeta^{(0)}_{H}(t) T_{H}(t)=0.
\end{equation}
This is the so-called homogeneous cooling state (HCS) \cite{Ha83}.
This state is known to be unstable with respect to long wavelength
spatial perturbations. More precisely, linear stability analysis of
the hydrodynamic equations shows that, for large enough systems, the
transversal component of the velocity relative to the square root of
the temperature grows in time leading to the instability of the
system \cite{GyZ93,McyY94,BDKyS98}. This shear instability has been
confirmed by molecular dynamics simulations \cite{GyZ93,McyY94} and
also by Monte Carlo simulations of the effective dynamics associated
to the Boltzmann equation \cite{BRyC96}. Of course, after a short
time interval, the theoretical predictions based on the linearized
hydrodynamic equations are not valid any longer since neglected
nonlinear effects become very important. During the initial stages
of the development of the instability, density inhomogeneities occur
in the system. For dilute systems whose linear size is larger (and
comparable) to the critical system size for the shear instability,
the spatial inhomogeneities seem to be dominated by a nonlinear
coupling of the density with the transversal velocity field
\cite{BRyC99}. Although this could indicate that the system is going
into a clustering regime, in which the particles tend to group
together forming very high density regions coexisting with very
dilute regions, simulation results indicate that the density
actually tends to a quite smooth profile \cite{BRyC99}.

Because of continuous cooling due to the inelasticity of collisions,
stationary states are not possible in freely evolving granular gases
(for instance, with periodic boundary conditions). Nevertheless, it
is still possible that the ``final state'' reached by the system
when the HCS is unstable, exhibits some scaling properties so that
it can be simply identified at least at the hydrodynamic level of
description. Here, attention will be focused on systems just above
the shear mode instability threshold. {\em Well} above this
threshold, the physical scenario might be quite different
\cite{ELyM05,MFyV08,PAFyM08}.

Consider hydrodynamic flows verifying the following conditions: i)
there are gradients in only one direction taken as the $X$ axis, and
ii) the hydrodynamic velocity field has the form $u_{i}({\bm r},t)=
\delta_{i,y} u(x,t)$. Other specifications of the state of the
system will be made when appropriate. Use of the above conditions
into Eqs.\ (\ref{2.1})-(\ref{2.3}) shows that the density profile
must be stationary, $n=n(x)$, and the component ${\sf P}_{xx}$ of
the pressure tensor must be uniform. In the framework of the
Navier-Stokes approximation, the latter implies that the pressure is
also uniform, $p=p(t)$. Moreover, the equations for the velocity and
temperature fields become
\begin{equation}
\label{2.12} \frac{\partial u}{\partial t}- (mn)^{-1}
\frac{\partial}{\partial x} \left( \eta \frac{\partial u}{\partial
x}\right)=0,
\end{equation}
\begin{equation}
\label{2.13} \frac{\partial T}{\partial t}-2 (n d)^{-1} \left\{ \eta
\left( \frac{\partial u}{\partial x} \right)^{2} +
\frac{\partial}{\partial x} \left[ \left( \kappa -\frac{n \mu}{T}
\right) \frac{\partial T}{\partial x} \right] \right\} + T
\zeta^{(0)} =0.
\end{equation}
To solve the above equations, the boundary conditions must be
specified. A system of $N$ particles enclosed in a cubic ($d=3$) or
square ($d=2$) box of side $L$ will be considered. Periodic boundary
conditions will be assumed at all the boundaries in order to avoid
undesirable wall effects. It will be convenient for the purposes
here to introduce a function $\varphi (x)$ by
\begin{equation}
\label{2.14} n(x)= n_{H} \left[ 1 + \varphi (x) \right],
\end{equation}
with $n_{H} \equiv N/L^{d}$. The conservation of the number of
particles and the periodic boundary conditions imply that
\begin{eqnarray}
\label{2.15} \int_{0}^{L} dx\, \varphi (x) & = &0, \nonumber \\
\varphi (x+L) & = &\varphi (x), \quad  0 \leq x <L.
\end{eqnarray}
Moreover, it will be assumed that $| \varphi (x) | \ll 1$, i.e. the
spatial density inhomogeneities are supposed to be small. It will be
shown in the following that this actually reduces the range of
applicability of the theory to systems near (and above) the
threshold of the shear instability. Then, the temperature profile in
the state being considered is given by
\begin{equation}
\label{2.16} T(x,t) \approx \theta (t) \left[ 1 - \varphi (x)
\right]
\end{equation}
where
\begin{equation}
\label{2.17} \theta (t) \equiv \frac{p(t)}{n_{H}}=
\frac{\int_{0}^{L}dx\, n(x) T(x,t)}{\int_{0}^{L} dx\, n(x)}
\end{equation}
is the spatial average temperature of the system at time $t$. It
must be realized that assuming $|\varphi | \ll 1$ and, consistently,
retaining terms up to first order in it when solving Eqs.\
(\ref{2.12}) and (\ref{2.13}), is not the same as linearizing around
the HCS, since nothing is in principle assumed about the amplitude
of the temperature $\theta(t)$ or the velocity field $u(x,t)$. This
will be discussed in detail in the next section.

\section{The free shear state}
\label{s3} The hydrodynamic equations (\ref{2.12}) and (\ref{2.13})
can be simplified by using dimensionless length and time scales
defined by
\begin{equation}
\label{3.1} l \equiv \frac{\nu (t)}{2} \left[ \frac{m}{\theta(t)}
\right]^{1/2} x,
\end{equation}
and
\begin{equation}
\label{3.2} s \equiv \frac{1}{2} \int_{0}^{t} dt^{\prime}\, \nu
(t^{\prime}),
\end{equation}
respectively. Here
\begin{equation}
\label{3.3} \nu(t) \equiv \frac{n_{H} \theta
(t)}{\eta_{0}(\theta)}\,
\end{equation}
is a characteristic frequency, proportional to the Boltzmann
collision frequency $\nu_{B}(t)$, namely $\nu_{B}(t) = (2+d) \nu (t)
/4$. Note that the pre-factor in Eq.\ (\ref{3.1}) does not depend on
time and the length scaling is made with a characteristic length
which is proportional to the (time-independent) average mean free
path of the system. In terms of the new scales, Eq. (\ref{2.12}) can
be approximated by
\begin{equation}
\label{3.4} \frac{\partial u(l,s)}{\partial s}-
\frac{\eta^{*}(\alpha)}{2} \frac{\partial^{2} u(l,s)}{\partial
l^{2}}=0,
\end{equation}
where terms of order $u \varphi$ have been neglected. The solution
of the above equation is a superposition of monocromatic waves of
the form
\begin{equation}
\label{3.5} u_{k}(l,s)= \chi_{k} (l) \omega_{k}(s)
\end{equation}
with
\begin{equation}
\label{3.6} \chi_{k}(l)=\sin (kl+\phi_{k})
\end{equation}
and
\begin{equation}
\label{3.7} \omega_{k} (s) = \omega_{k,0} e^{- \frac{k^{2} \eta^{*}
s}{2}},
\end{equation}
where $\phi_{k}$  and $\omega_{k,0}$ are arbitrary constants. The
possible values of $k$ are restricted by the periodic boundary
conditions which imply
\begin{equation}
\label{3.8} k=\frac{2 \pi q}{l_{M}},
\end{equation}
$q$ being a positive integer and $l_{M}$ the value of $l$ for $x=L$,
i.e.
\begin{equation}
\label{3.9} l_{M}= \frac{\nu(t)}{2}\left[ \frac{m}{\theta (t)}
\right]^{1/2}L.
\end{equation}
The form of Eq. (\ref{3.7}) indicates that in the limit of large
time $s$, the dynamics of the system is governed by the fundamental
mode corresponding to the lowest possible value of $k$, i.e.
$k=k_{m}= 2 \pi /l_{M}$. Therefore, for large enough times,
\begin{equation}
\label{3.10} u(l,s) \approx  u_{k_{m}}(l,s) = \chi (l) \omega(s).
\end{equation}
with $\chi (l) = \chi_{k_{m}}(l)$ and $\omega(s) =
\omega_{k_{m}}(s)$. In the same approximation under consideration,
$|\varphi (x)| \ll 1$, the evolution equation for the temperature
(\ref{2.13}) leads to
\begin{equation}
\label{3.11} \frac{1}{\theta (s)}\frac{d \theta (s)}{d s} - \frac{m
\eta^{*}(\alpha) \omega^{2}(s)}{\theta (s)d} \left(
\frac{d\chi(l)}{dl} \right)^{2}+ \zeta^{*} (\alpha)[ 2 + \varphi
(l)]+\frac{d+2}{2(d-1)} \left[ \kappa^{*}(\alpha)-\mu^{*}(\alpha)
\right] \frac{d^{2}\varphi (l)}{d l^{2}} =0.
\end{equation}
Using Eq.\, (\ref{3.6}) this becomes
\begin{eqnarray}
\label{3.12} \frac{1}{\theta (s)}\frac{d \theta (s)}{d s}-
\frac{\eta^{*}(\alpha)m \omega^{2}(s)k_{m}^{2}}{2\theta (s) d}
\left\{ 1+ \cos [ 2(k_{m} l + \phi_{k_{m}} )] \right\}  & &
\nonumber
\\ + \zeta^{*} (\alpha)[2 +\varphi (l)]+\frac{d+2}{2(d-1)} \left[
\kappa^{*}(\alpha)-\mu^{*}(\alpha) \right] \frac{d^{2}\varphi (l)}{d
l^{2}} =0. &&
\end{eqnarray}
By requiring the sum of the position dependent terms to cancel it is
obtained that
\begin{equation}
\label{3.13} \varphi (l)= - \frac{(d-1) \eta^{*} (\alpha)}{4 d(d+2)
\gamma (k_{m}, \alpha)}\, a \cos[2(k_{m}l +\phi_{k_{m}})],
\end{equation}
where
\begin{equation}
\label{3.12a} \gamma(\alpha,k_{m}) \equiv \kappa^{*}(\alpha)-\mu^{*}
(\alpha) -\frac{(d-1) \zeta^{*}(\alpha)}{2(d+2) k_{m}^{2}}
\end{equation}
and
\begin{equation}
\label{3.14} a \equiv \frac{m \omega^{2}(s)}{\theta (s)}\, .
\end{equation}
Consistency requires that $a$ be actually independent of $s$ and,
moreover, that it be defined positive. The former of these
conditions together with Eq. (\ref{3.4}) leads to
\begin{equation}
\label{3.15} \frac{1}{\theta (s)}\frac{d \theta (s)}{d s} =
\frac{2}{\omega (s)} \frac{d \omega (s)}{d s} =- k_{m}^{2}
\eta^{*}(\alpha).
\end{equation}
This equation will be used later on to determine the effective
cooling rate of the average temperature $\theta (s)$ of the system
in the free shear state. On the other hand, equating to zero the
part of Eq. (\ref{3.12}) that is position independent yields, after
employing Eq. (\ref{3.15}),
\begin{equation}
\label{3.16} a= \frac{ 2 d [2 \zeta^{*}(\alpha)- k_{m}^{2}
\eta^{*}(\alpha)]}{k_{m}^{2} \eta^{*} (\alpha)}\, .
\end{equation}
As mentioned above this quantity {\em must} be positive. As a
consequence, the existence of the mathematical solution of the
hydrodynamic Navier-Stokes equations under consideration and hence
of the free shear state is only posible if
\begin{equation}
\label{3.17} k_{m}^{2} < \frac{2 \zeta^{*}}{\eta^{*}},
\end{equation}
or, equivalently, $L>L_{c}$, with the critical size $L_{c}$ given by
\begin{equation}
\label{3.18} L_{c}=\frac{(2+d) \Gamma \left( d/2 \right)}{2
\pi^{\frac{d-3}{2}} n_{H} \sigma^{(d-1)}}\, \left( \frac{\eta^{*}}{2
\zeta^{*}} \right)^{1/2}.
\end{equation}
This coincides with the condition determining the instability region
of the shear mode found in the linear stability analysis of the
hydrodynamic equations around the HCS \cite{BDKyS98}.

Substitution of Eq. (\ref{3.16}) into Eq. (\ref{3.13}) provides the
explicit form of the steady density profile in the free shear state
near the threshold of the instability,
\begin{equation}
\label{3.19} \varphi (l) = - A \cos[2 (k_{m}l +\phi_{k_{m}})],
\end{equation}
\begin{equation}
\label{3.20} A= \frac{(d-1) \eta^{*}(\alpha)}{2(d+2) \gamma
(k_{m},\alpha)} \left[  \left(\frac{L}{L_{c}} \right)^{2} -1
\right].
\end{equation}
Therefore, the condition $|\varphi (x)| \ll 1$  formally implies
that $|L-L_{c}|/L_{c} \ll 1$. How restrictive this condition
actually  is will be seen in the next section.

A particularly simple expression for the amplitude of the velocity
field is obtained by scaling with the average temperature. Equations
(\ref{3.14}) and (\ref{3.16}) yield
\begin{equation}
\label{3.21} \widetilde{\omega} (s) \equiv \omega(s) \left[
\frac{m}{ 2 \theta (s)} \right]^{1/2} =d^{1/2} \left[
\left(\frac{L}{L_{c}} \right)^{2} -1 \right]^{1/2}.
\end{equation}
Therefore, the scaled macroscopic velocity field is time-independent
and all its dependence on the average density and inelasticity
occurs through the critical length $L_{c}$. A similar expression can
be obtained for the ratio between the effective cooling rate of the
shear state $\zeta_{S}^{*}$ and the cooling rate of the HCS,
$\zeta^{*}$. The former is identified by writing Eq. (\ref{3.15}) in
the form
\begin{equation}
\label{3.22} \frac{\partial \theta (s)}{\partial s} + 2
\zeta^{*}_{S} \theta (s) =0,
\end{equation}
with
\begin{equation}
\label{3.23} \zeta_{S}^{*} =\frac{k_{m}^{2}\eta^{*}(\alpha)}{2},
\end{equation}
that leads to
\begin{equation}
\label{3.24} \frac{\zeta_{S}^{*}}{\zeta^{*}} = \left(
\frac{L_{c}}{L} \right)^{2}.
\end{equation}
Of course, in the limit $L \rightarrow L_{c}$, $\theta (s)
\rightarrow T_{H}(s)$ and the law for the HCS given by Eq. (11) is
recovered.

The average energy per particle is
\begin{equation}
\label{3.25} \overline{e}_{T} \equiv \frac{m}{2}\, \overline{u^{2}}
+ \frac{\overline{T}d}{2},
\end{equation}
with the bar denoting spatial average. Using Eqs. (\ref{2.17}),
(\ref{3.5}), (\ref{3.6}), (\ref{3.14}), and (\ref{3.16}) it is
found:
\begin{equation}
\label{3.26} \overline{e}_{T} (s)= \frac{\zeta^{*} (\alpha) \theta
(s) d}{k_{m}^{2} \eta^{*}(\alpha)}= \left( \frac{L}L_{c} \right)^{2}
\frac{\theta (s)d}{2}\, .
\end{equation}
For $L \rightarrow L_{c}$, the equilibrium relation
$\overline{e}_{T} = d T /2$ is recovered as expected. This equation
has been previously obtained by Wakou { \em et al.} \cite{WByE02},
in the context of a Landau-Ginzburg-type equation of motion derived
from the hydrodynamic equations, under certain restrictions and in
the limit of nearly elastic collisions ($1-\alpha \ll 1$). At this
point, it is worth mentioning that the assumed periodic boundary
conditions do not play an essential role in the theory developed
here, although they must be compatible with the symmetry of the
shear state. For instance, it is easily realized that completely
equivalent results are obtained if elastic walls were used at the
boundaries of the system.

Soto {\em et al.,} \cite{SMyM00} have carried out a nonlinear
analysis of the hydrodynamic equations of a system of inelastic hard
disks close to the instability threshold. When $k_{m}$ is slightly
smaller than $k_{c} \equiv 2 \pi /L_{c}$, vorticity modes with
wavenumber $k_{m}$ exhibit a critical slowing down, so that all the
other hydrodynamic modes can be considered as enslaved by them.
Then, it is possible to derive closed equations for the amplitudes
of the vorticity modes with $k=k_{m}$ by using the adiabatic
elimination method. This approach can be expected to be formally
consistent with the one presented here, but quantitative and
qualitative discrepancies occur when comparing the results from both
approaches. Although the authors of \cite{SMyM00} do not give almost
any detail of their analysis of the nonlinear hydrodynamic
equations, we have identified a twofold origin of the discrepancies.
Firstly, a nonlinear term involving the vorticity seems to have been
inconsistently neglected. Secondly, the linearization around the HCS
and the adiabatic method used in \cite{SMyM00} is not equivalent to
the approximation scheme presented here, in which the velocity field
obeys the closed Eq.\ (\ref{3.4}).

\section{Direct Monte Carlo simulations }
\label{s4} To test the theoretical predictions presented in the
previous sections and, in particular, the existence itself of the
free shear state, we have employed the direct simulation Monte Carlo
(DSMC) method \cite{Bi94,Ga00}, to numerically generate the dynamics
of a system of inelastic hard disks. The DSMC method is a
many-particle algorithm designed to mimic the effective dynamics of
the particles of a gas in the low density limit. Therefore, in this
limit it is expected to lead to the same results as, for instance,
molecular dynamics simulations, with the advantage of a much larger
statistical accuracy. In all the simulations, the initial state was
homogeneous and isotropic with a Gaussian velocity distribution, and
a square box with periodic boundary conditions was used. The linear
size of the system $L$ was always larger than, but close to, the
critical one $L_{c}$, predicted by Eq.\ (\ref{3.18}), for the
considered value of the restitution coefficient $\alpha$. Then,
according to the theory, the system is expected to generate the
non-linear free shear flow described in Sec. \ref{s3}, if it is
stable.

As usual in DSMC simulations, the mass $m$ of the particles will be
taken as the unit of mass, the average mean free path $\lambda
\equiv (2 \sqrt{2} n_{H} \sigma)^{-1}$ as the unit of length, and $2
T(0) \equiv 2 \theta (0)$, where $T(0)$ is the initial temperature,
as the unit of energy. The effect of a collision between particles
$r$ and $s$ is to instantaneously modify their velocities according
to the rule
\begin{equation}
\label{4.1} {\bm v}_{r} \rightarrow {\bm v}^{\prime}_{r} = {\bm
v}_{r} -\frac{1+\alpha}{2} (\widehat{\bm \sigma} \cdot {\bm v}_{rs})
\widehat{\bm \sigma},
\end{equation}
\begin{equation}
\label{4.2} {\bm v}_{s} \rightarrow {\bm v}^{\prime}_{s} = {\bm
v}_{s} +\frac{1+\alpha}{2} (\widehat{\bm \sigma} \cdot {\bm v}_{rs})
\widehat{\bm \sigma},
\end{equation}
where ${\bm v}_{rs} \equiv {\bm v}_{r} -{\bm v}_{s}$ is the relative
velocity and $\widehat{\bm \sigma}$ is the unit vector pointing from
the center of particle $s$ to the center of particle $r$ at contact.
This corresponds to the scenario in which the constitutive relations
(\ref{2.4})-(\ref{2.6}) were derived \cite{BDKyS98,ByC01}.

One of the technical difficulties when numerically simulating a
freely evolving granular fluid is the continuous  cooling of the
system, i.e. the decrease in magnitude of the typical velocity of
the particles. As a consequence, the numerical inaccuracies become
very large after some time interval. To avoid this difficulty, a
procedure was introduced  based on a change in the time scale being
used. More specifically, a logarithmic time scale is introduced
\cite{Lu01,BRyM04}. Then, the dynamics of systems in states where
all the time dependence comes through the average temperature
because of inelastic cooling, can be easily mapped into the dynamics
around steady states. In this case, the logarithmic time scale is
proportional to the cumulated number of collisions per particle.

\subsection{Transient dynamics}
\begin{figure}
\begin{tabular}{cc}
\includegraphics[scale=0.5,angle=0]{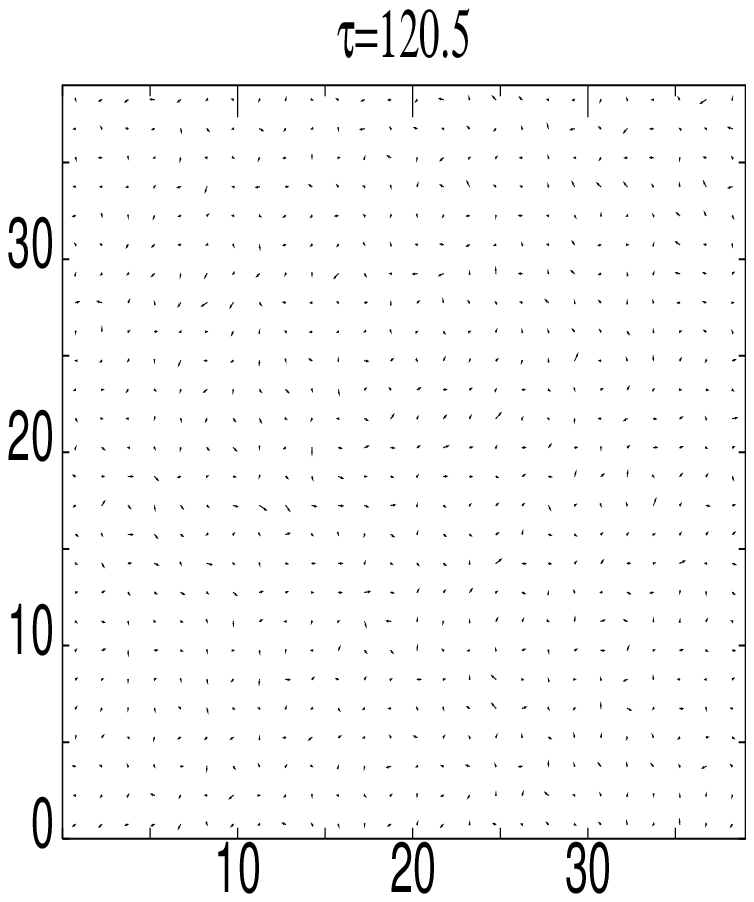} &
\includegraphics[scale=0.5,angle=0]{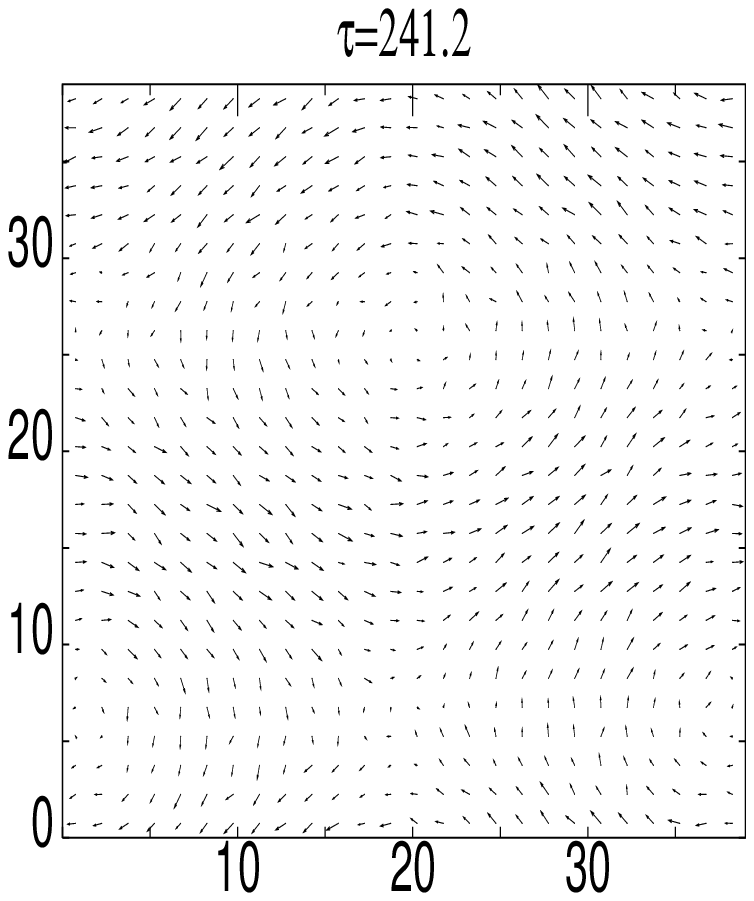} \\
\includegraphics[scale=0.5,angle=0]{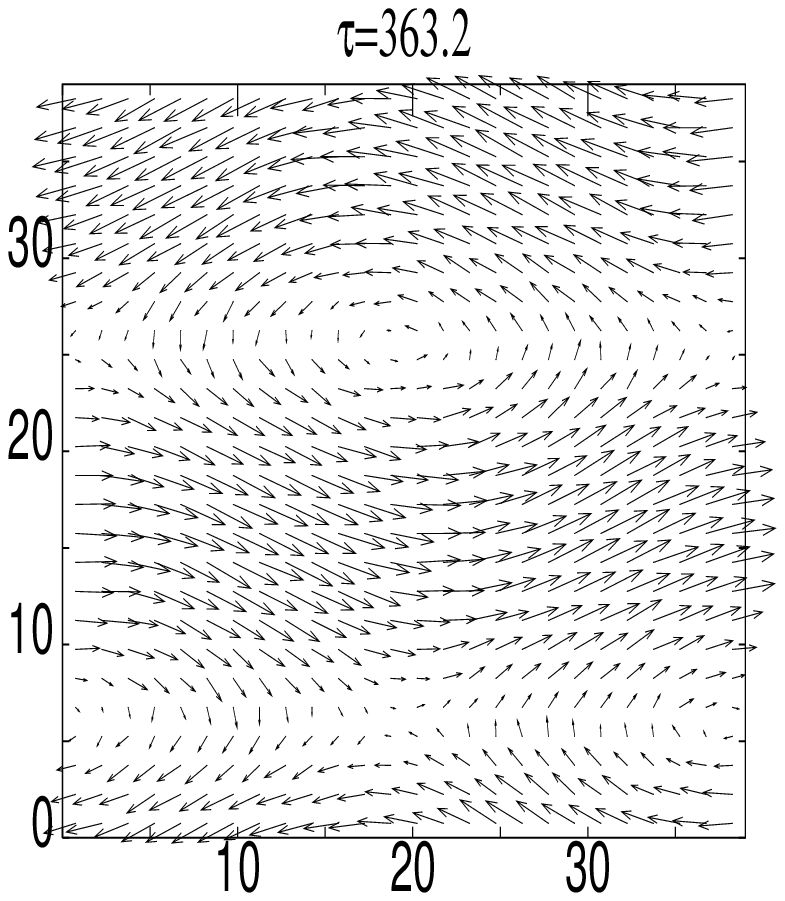} &
\includegraphics[scale=0.5,angle=0]{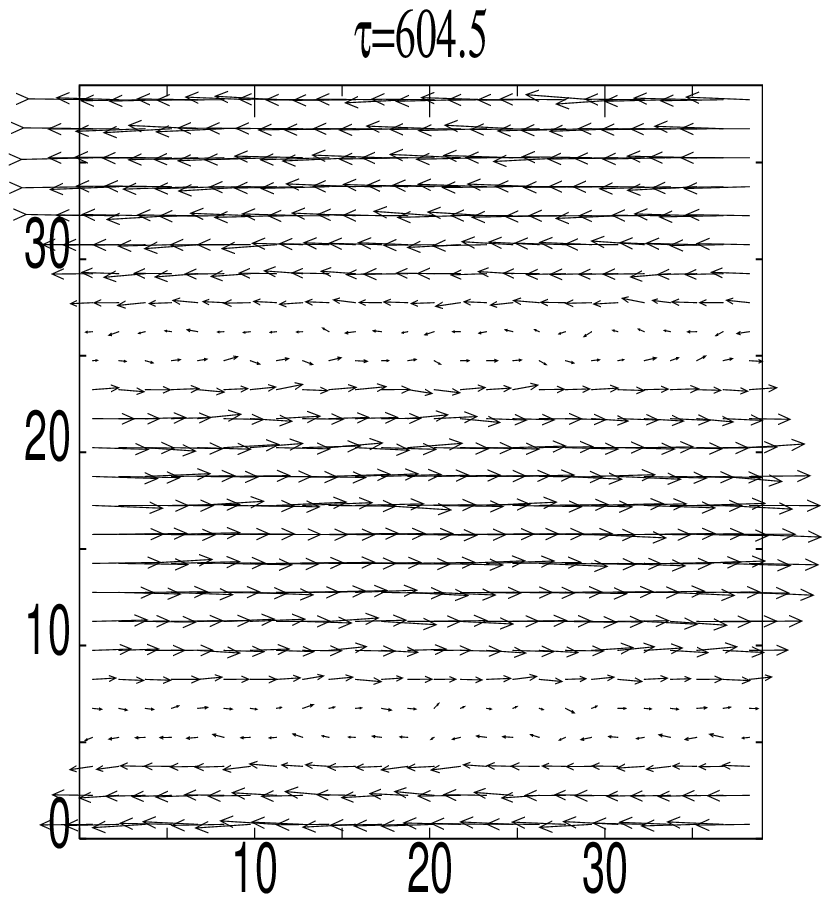}
\end{tabular}
\caption{Time evolution of the velocity field scaled with the
average thermal velocity, ${\bm u} (m/2 \theta)^{1/2}$, in a system
with $\alpha =0.95$ and $L/L_{c}\approx 1.205$. The indicated times
$\tau$ are measured as the average number of collisions per
particle. The lengths of the arrows in the plots are proportional to
the scaled velocities. \label{fig1}}
\end{figure}

In most of  the simulations to be reported, a similar time sequence
was observed. The system remains homogeneous with no macroscopic
velocity field for an initial period of time, developing afterwards
a state with two vortices, and finally a shear state characterized
by two counterflows parallel to one of the sides of the system. In
this state, there is a coupling between the velocity and density
fields. The spatial inhomogeneities remain stationary and the state
looks time independent in the logarithmic time scale used in the
numerical simulations. The particular location of the transient
vortices and their direction, and hence of the bands in the shear
state, depends on the initial state or, in a statistical sense, on
the fluctuation taking the system away from the HCS. An example of
the described behavior is given in Fig.\ \ref{fig1}, where the
scaled velocity field ${\bm u} \left(  m/ 2 \theta \right)^{1/2}$ is
plotted at four different times for a system with $\alpha =0.95$ and
$L=39$ ($L/L_{c} \approx 1.205)$. The formation of the vortices,
their distortion, and the formation of shear bands are clearly
identified. The latter remain unchanged after their formation, and
this was observed in all cases, at least as long as the system is
close enough to the shear instability.  The spontaneous symmetry
breaking occurring in the final shear state was observed in both
perpendicular directions and, to compare  with the theoretical
predictions, the $x$-axis was always chosen perpendicular to the
velocity field. More will be commented on this issue in the final
section of the paper.

\begin{figure}
\includegraphics[scale=0.5,angle=0]{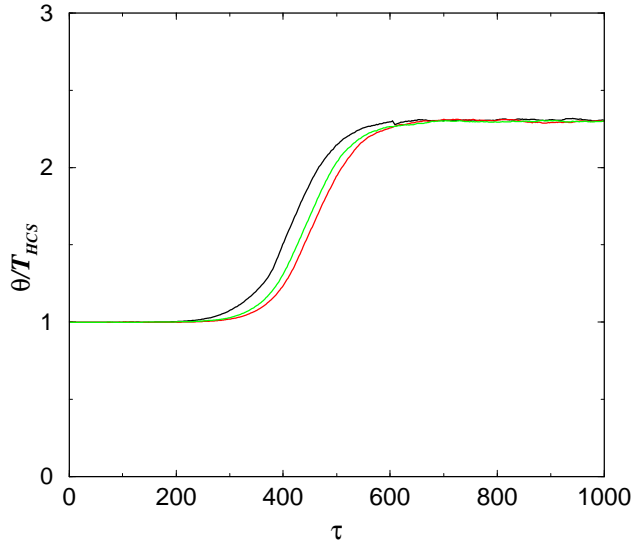}
\caption{(Color online) Time evolution of the spatial average
temperature normalized by the temperature of the reference HCS for
the same system as in Fig. \protect{\ref{fig1}}. Time $\tau$ is
measured as the average number of collisions per particle. The
several curves correspond to different simulation trajectories.
\label{fig2}}
\end{figure}

To get some additional insight about the time evolution of the
system before reaching the free shear state, in Fig.\ \ref{fig2} the
ratio between the spatial average temperature $\theta (t)$ and the
HCS temperature at the same time as predicted by the Haff law, Eq.
(\ref{2.11}), is shown as a function of the average number of
collisions per particle $\tau$. The system is the same as in Fig.
\ref{fig1} and three different simulation trajectories are plotted.
Although the time evolution of the temperature is not exactly the
same in all cases, a quite similar trend is observed. There is a
time interval in which the average temperature is well described by
the Haff law ($\tau \lesssim 250$) in spite of the system having
already well developed vortices (see Fig. \ref{fig1}). Afterwards,
for $ 250 \lesssim \tau \lesssim 600$, the average temperature
decays much slower than in the HCS, so that the ratio grows very
fast, until it saturates at a given value. The region with the
largest growth corresponds to the distortion of the velocity
vortices. The constant value reached in the final shear state
indicates that the average temperature $\theta (t)$ still obeys a
Haff-like law, but with a different cooling rate, in qualitative
agreement with Eq.\ (\ref{3.22}).

\begin{figure}
\includegraphics[scale=0.5,angle=0]{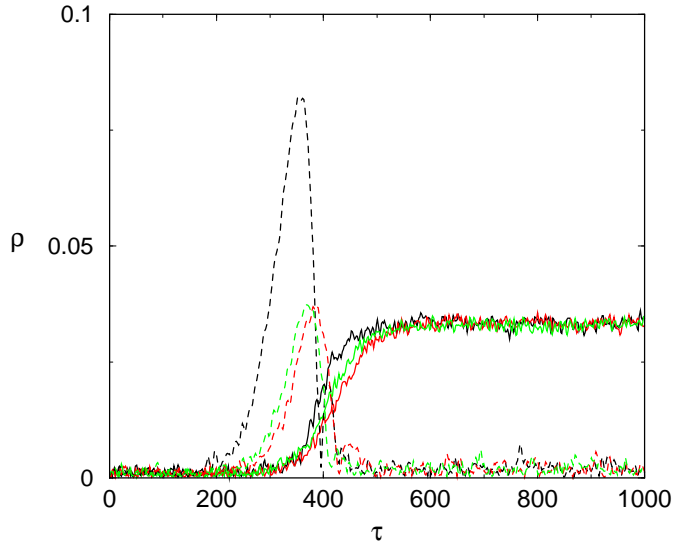}
\caption{(Color online) Time evolution of the second transversal
mode of the relative density $\rho_{2k_{m}}$ (solid lines) and the
diagonal mode $\rho_{k_{m},k_{m}}$ (dashed lines). The system is the
same as in Fig. \protect{\ref{fig1}}. The several curves correspond
to different trajectories, and time $\tau$ is measured again as the
average number of collisions per particle. \label{fig3}}
\end{figure}

The evolution of the density is illustrated in Fig.\ \ref{fig3},
also for the system with $\alpha =0.95$ and $L/L_{c} \approx 1.205$.
Two Fourier components of the relative density field $\rho \equiv
n/n_{H}$ are plotted, $\rho_{2k_{m}}$ and $\rho_{k_{m},k_{m}}$. The
former corresponds to the second density mode that is the one
predicted to survive by the theory, Eq.\ (\ref{3.19}). In the
simulations, it appears parallel to any of the sides of the system
as discussed above and no distinction is made when reporting the
simulation results. The other component, $\rho_{k_{m},k_{m}}$ is the
lowest mode along the diagonal of the system. Again, the several
curves are different simulation trajectories. In all cases, the
second transversal mode of the density grows from the noise level to
a constant value as the shear instability develops. Moreover, the
final steady  value is the same for all trajectories. On the other
hand, although the diagonal mode often also grows in the time
interval in which the two vortices are losing their shape
transforming into the shear state, it decays to the noise level once
this state is reached.

\subsection{Shear state results}
\begin{figure}
\includegraphics[scale=0.5,angle=0]{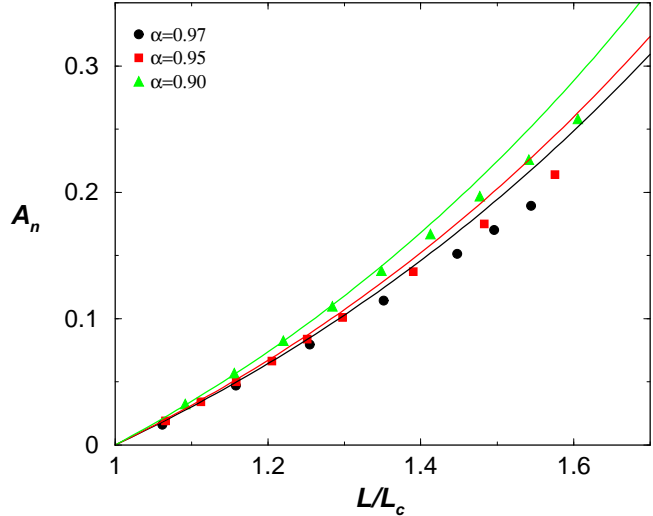}
\caption{(Color online) Dimensionless amplitude of the cosine
density profile in the free shear state as a function of the ratio
between the size of the system $L$ and its critical value for the
shear instability $L_{c}$. Three values of the restitution
coefficient are considered, as indicated. The symbols are simulation
results and the lines are the predictions of Eq.\
(\protect{\ref{3.20}}), the lowest line corresponding to the larger
$\alpha$.\label{fig4}}
\end{figure}

Now the results obtained for the properties of the system once in
the free shear state will be reported, and compared with the
theoretical predictions obtained in this paper. Consider first the
density profile. The simulation data are very well fitted by a
cosine function as given in Eq. (\ref{3.19}). This has been checked
both by plotting directly the profiles and by computing their
Fourier components. The amplitude of the measured perturbation
$A_{n}$ is plotted in Fig.\ \ref{fig4} as a function of $L/L_{c}$
for three values of the restitution coefficient, $\alpha = 0.97,
0.95$, and $0.9$. The lines are the theoretical predictions given by
Eq.\ (\ref{3.20}), the highest one corresponding to the smallest
value of $\alpha$ and the lowest one to the greatest value of
$\alpha$. It is seen that the agreement is quite good when $L/L_{c}$
is close to $1$, the discrepancies increasing as the size of the
system increases above its critical value. Also, the deviation is
larger the smaller the coefficient of restitution. In any case, it
must be kept in mind that the theory developed here is by
construction restricted to states with $A_{n} \ll 1$.

\begin{figure}
\includegraphics[scale=0.5,angle=0]{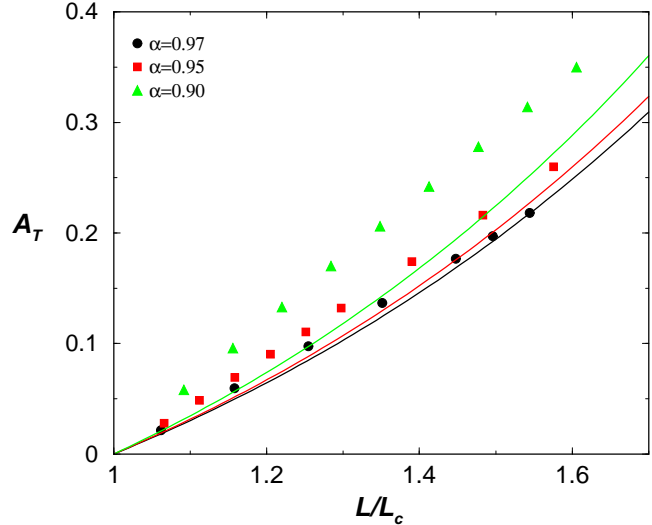}
\caption{(Color online) The same as in Fig.\ \protect{\ref{fig4}}
but for the amplitude of the cosine temperature
profile.\label{fig5}}
\end{figure}

In Fig. \ref{fig5}, the amplitude $A_{T}$ of the steady cosine
profile for $1-T(x,t)/\theta(t)$ is presented for the same values of
the parameters as considered in Fig.\ \ref{fig4}. According to the
theory developed here, Eq.\ (\ref{2.16}), this amplitude should be
the same as for the density profile, i.e. $A_{T}=A_{n}=A$.
Nevertheless, the simulation data indicate that this is not the
case, except for values of  $L/L_{c}$ close to unity. The deviations
of the numerical data from the theoretical predictions are in
opposite directions for $A_{T}$ and $A_{n}$, being larger for the
former. This indicates that the pressure is not actually strictly
uniform as predicted by the Navier-Stokes equations for dilute
granular gases, but exhibits some oscillatory profile. On the other
hand, the component ${\sf P}_{xx}$ of the pressure tensor was found
to be uniform, as required by the balance equations
(\ref{2.1})-(\ref{2.3}). Nevertheless, for $L/L_{c}$ small enough,
the agreement between theory and simulation can be qualified as
satisfactory. Let us mention that if the elastic values of the
transport coefficients ($\eta^{*}=\kappa^{*}=1, \mu^{*}=0$) were
used, the theoretical prediction for $A$ as a function of $L/L_{c}$
would become independent of $\alpha$ contrary to what is observed in
the simulations.

\begin{figure}
\includegraphics[scale=0.5,angle=0]{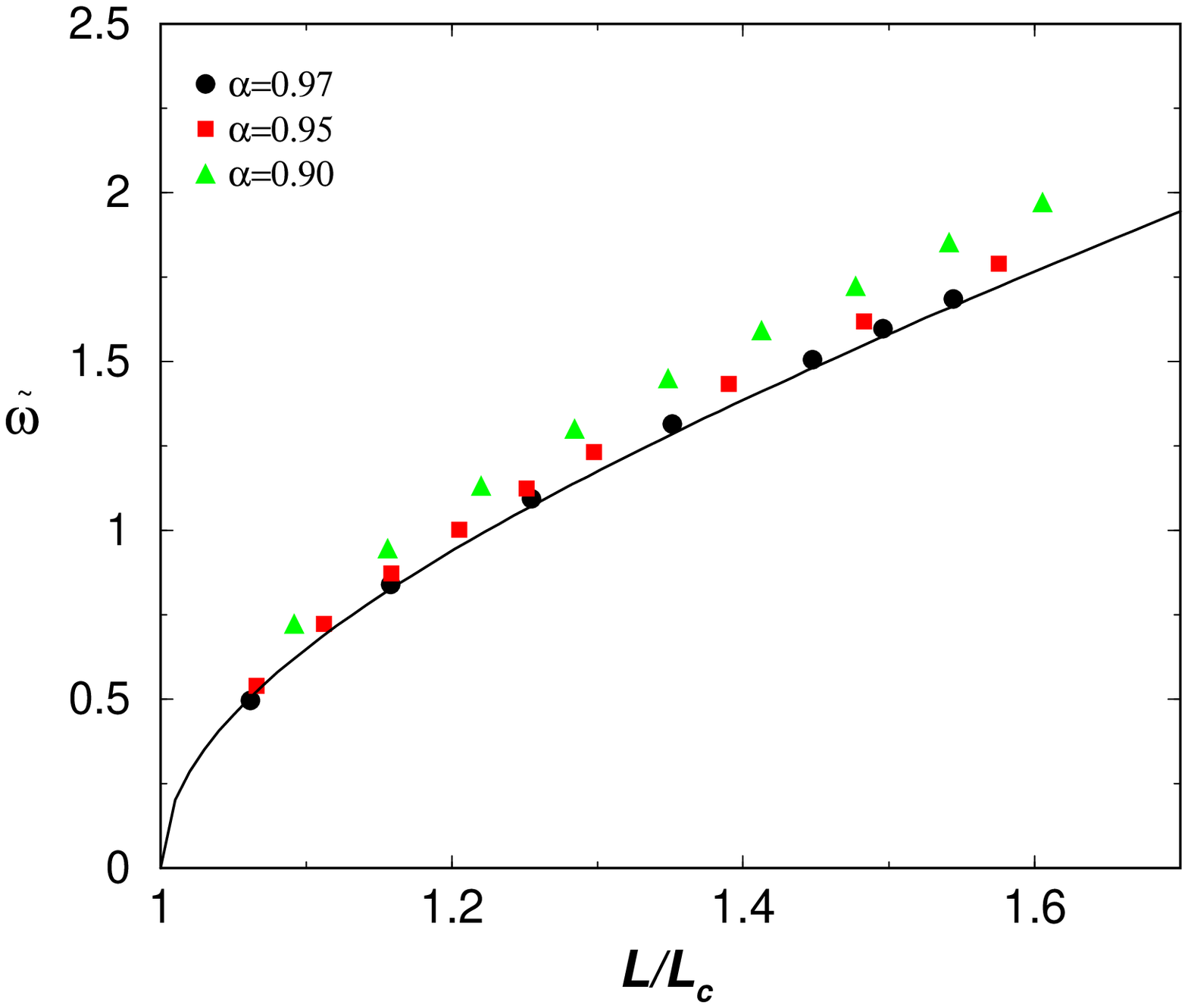}
\caption{(Color online) Dimensionless amplitude of the macroscopic
flow $\widetilde{\omega} \equiv \omega(t) / \left(2 \theta(t)/m
\right)^{1/2}$ in the free shear state as a function of the length
$L$ of the system normalized by its critical value $L_{c}$. The
solid line is the theoretical prediction, Eq.\
(\protect{\ref{3.21}}), while the symbols are simulation results for
the same systems as in Fig.\ \protect{\ref{fig4}}.\label{fig6}}
\end{figure}

A quite strong prediction of the theory is provided by Eq.\
(\ref{3.21}), where the amplitude of the velocity field scaled with
the square root of the average granular temperature is expressed as
a simple time and $\alpha$-independent function of the ratio
$L/L_{c}$. This latter property was seen to ve verified in the
simulation within the statistical uncertainties. Moreover, the
results displayed in Fig.\ \ref{fig6} show that the simulation data
are in good agreement with the theoretical expression, although
again systematic deviations are observed as the value of the
coefficient of restitution decreases and/or the size of the system
as compared with its critical value increases.

\begin{figure}
\includegraphics[scale=0.5,angle=0]{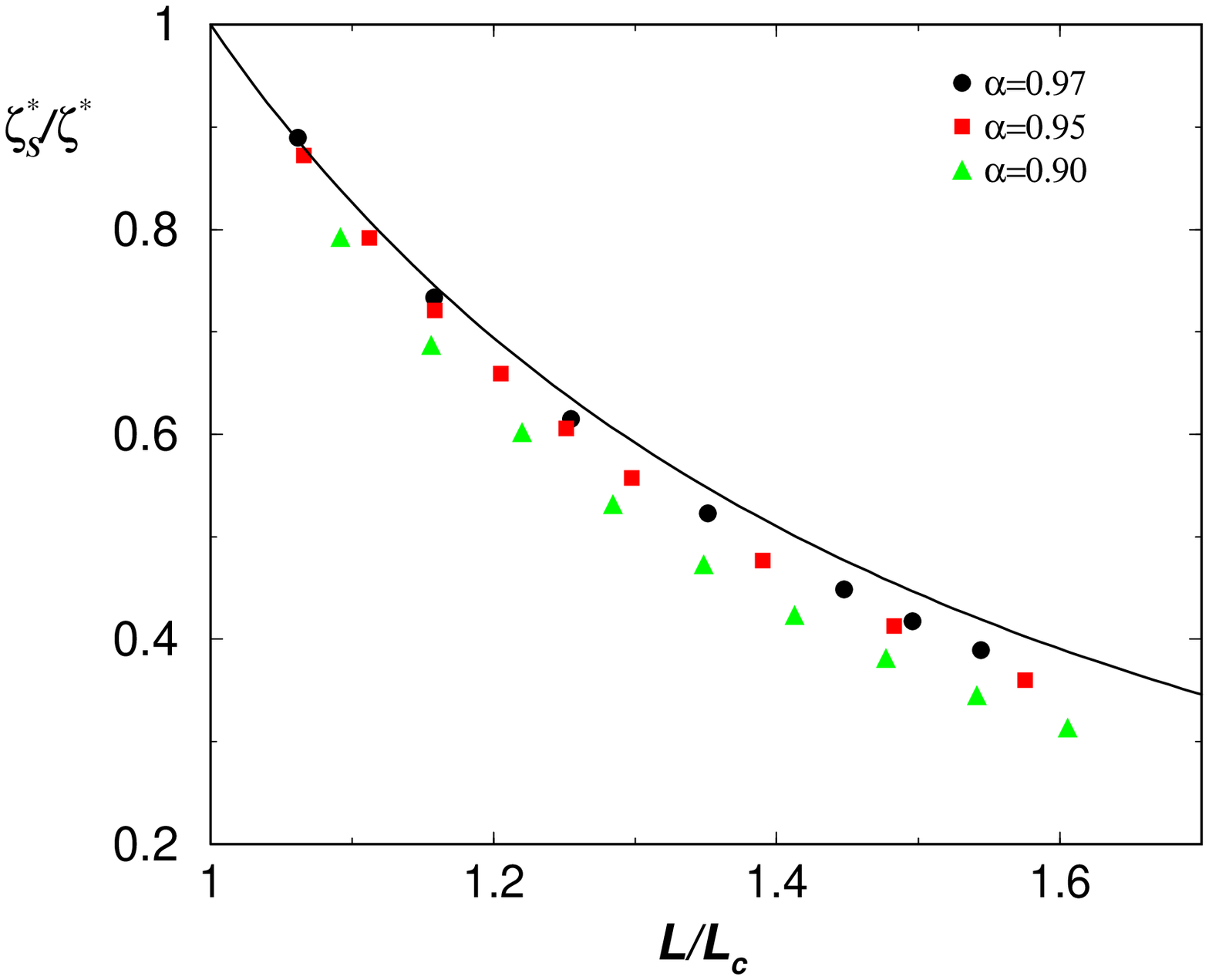}
\caption{Ratio between the cooling rates of the free shear state and
of the HCS as a function of the ratio $L/L_{c}$. The symbols have
the same meaning as in Fig.\ \protect{\ref{fig4}} and the solid line
is the theoretical prediction, Eq.\, (\protect{\ref{3.23}}).
\label{fig7}}
\end{figure}

The simulation results also indicate that the decay of the average
temperature in the free shear state is governed by a Haff-like law,
Eq. (\ref{3.21}).  In Fig.\ \ref{fig7}, the ratio between the
cooling rate of the free shear state, $\zeta_{S}^{*}$, and the one
of the HCS, $\zeta^{*}$, is plotted, for the same systems being
considered along this section. The theoretical prediction is
provided by Eq.\ (\ref{3.24}), implying that the free shear state
cools slower than the associated HCS, i.e. with the same density and
initial temperature. Again a quite satisfactory agreement is found
between theory and simulations, with the discrepancies exhibiting
the same trends as in all the previous comparisons.

\section{Summary and discussion}
\label{s5}

It has been shown that the hydrodynamic Navier-Stokes equations for
granular gases predict the existence of a shear state for  freely
evolving systems, whose size is slightly larger than the critical
size for the shear mode instability. Explicit analytical expressions
for the hydrodynamic fields have been derived. Although it is a
non-stationary state due to cooling, all the time dependence of the
hydrodynamic fields  occurs through the average temperature and,
therefore, it can be eliminated by introducing appropriate
dimensionless quantities. These theoretical predictions have been
found to be in good agreement with the numerical results obtained by
the DSMC method, for values of the restitution coefficient $\alpha$
close to unity. When the system is more inelastic, significant
deviations from the theory are observed. This was expected, since
the free shear state is characterized, as many other non-equilibrium
states of granular gases, by a strong coupling between gradients and
inelasticity. In the present case, this is easily identified through
the dependence on $\alpha$ of $L_{c}$. As a consequence, a first
order gradient expansion like the one leading to the Navier-Stokes
equations also implies a limitation in the range of values of
$\alpha$ for which the theory applies. Moreover, it is probably true
that the free shear state is inherently non-Newtonian, as it is the
case for the steady uniform shear state of a granular gas
\cite{SGyD04}. A possible indication of this is the inhomogeneity of
the pressure observed in the simulations as implied by the
difference between the measured values of $A_{n}$ and $A_{T}$.

In ref. \cite{PAFyM08} several possible hydrodynamic scenarios are
described for the behavior of a freely evolving granular fluid
inside its instability region. In that classification, the situation
considered in this paper is called scenario 4. Here  the {\em final}
state reached by the system has been investigated and a quantitative
test for the complete scenario has been provided. The free shear
state seems stable, in the sense that no deviations from it have
been observed in the simulations. In this context, it is worth
mentioning that, in many cases, the two-vortex  state was clearly
identified for a quite large period of time before the system moved
to the shear state. This may indicate that it is a metastable state
with a large escape time. This issue clearly deserves more
attention. On the other hand, when $L$ becomes much larger than
$L_{c}$ (typically, $L/L_{c} \agt 1.7$), other hydrodynamic modes
become relevant and the free shear state is not expected to occur in
the system. This has been confirmed by the simulation results.

It is also worth comparing in some detail the approach  followed
here with the nonlinear stability analysis carried out in ref.
\cite{SMyM00}. As already mentioned, the results reported there
suggest that a term nonlinear in the velocity  has been omitted
although it is of the same order as others that are kept. This term
dramatically modifies the results of the stability analysis
\cite{DMGByR08}. Moreover, in \cite{SMyM00} an adiabatic elimination
method is used, in which the time derivative of the hydrodynamic
fields other than the transverse velocity are set equal to zero. In
this way, these fields can be expressed in terms of $u$, and when
the expressions are inserted into the equation for $u$, a closed
equation is obtained for the latter. The approximation followed here
is different. To lowest order, the transversal velocity field obeys
a closed equation by itself, Eq.(\ref{3.4}), once the function
$\theta (t)$ has been scaled out by the change of variables. This
equation differs from the one obtained by the adiabatic elimination
method. It is not fully clear to us presently the physical origin of
this strong discrepancy.

One relevant question that always arises when using smooth inelastic
hard particles to model granular gases, is to what extent the
results would be modified if more realistic models, including for
instance rotational degrees of freedom \cite{GNyB05} and/or velocity
dependent restitution coefficients \cite{ByP04}, were considered.
Although the hydrodynamic Navier-Stokes equations including these
effects are known in some limiting cases, their analysis is far
beyond our present reach. Nevertheless, it can be expected that an
extension of the energy balance appearing in the simple case
discussed here will hold when more dissipation mechanisms are
included. Then it is our conjecture that a similar shear state but
including the new rotational effects will also show up.

\section{Acknowledgements}

This research was supported by the Ministerio de Educaci\'{o}n y
Ciencia (Spain) through Grant No. FIS2008-01339 (partially financed
by FEDER funds).

\appendix

\section{Navier-Stokes transport coefficients}
\label{ap1}

In this Appendix, the explicit expressions of the transport
coefficients and the cooling rate introduced in Eqs.
(\ref{2.4})-(\ref{2.10}) are given. The elastic shear viscosity and
heat conductivity are
\begin{equation}
\label{ap1.1} \eta_{0}=\frac{2+d}{8} \Gamma \left( \case{d}{2}
\right) \pi^{-\frac{d-1}{2}} \left( m T \right)^{1/2} \sigma
^{-(d-1)},
\end{equation}
\begin{equation}
\label{ap1.2} \kappa_{0}=\frac{d(d+2)^{2}}{16(d-1)} \Gamma \left(
\case{d}{2} \right) \pi^{-\frac{d-1}{2}}  \left( \frac{T}{m}
\right)^{1/2} \sigma^{-(d-1)},
\end{equation}
respectively. As usual in the context of granular fluids, the
Boltzmann constant has been set equal to unity. The factors
accounting for the dependence on the restitution coefficient are
given by
\begin{equation}
\label{ap1.2a} \eta^{*}(\alpha)= \left[ \nu^{*}_{1}(\alpha)
-\frac{\zeta^{*}(\alpha)}{2} \right]^{-1},
\end{equation}
\begin{equation}
\label{ap1.3}
\kappa^{*}(\alpha)=[\nu^{*}_{2}(\alpha)-\frac{2d}{d-1}\zeta^{*}(\alpha)]^{-1}
 [1+c^{*}(\alpha)],
\end{equation}
\begin{equation}
\label{ap1.4} \mu^{*}(\alpha)=2\zeta^{*}(\alpha)\left[
\kappa^{*}(\alpha)+ \frac{(d-1)c^{*}(\alpha)}{2d\zeta^{*}(\alpha)}
\right] \left[ \frac{2(d-1)}{d}
\nu^{*}_{2}(\alpha)-3\zeta^{*}(\alpha) \right]^{-1},
\end{equation}
\begin{equation}
\label{ap1.5} \zeta^{*}(\alpha)=\frac{2+d}{4d} (1-\alpha^{2}) \left[
1+ \frac{3 c^{*}(\alpha)}{32} \right].
\end{equation}
In the above expressions,
\begin{equation}
\label{ap1.5a} \nu^{*}_{1}(\alpha) = \frac{(3-3 \alpha
+2d)(1+\alpha)}{4d} \left[ 1- \frac{c^{*}(\alpha)}{64} \right],
\end{equation}
\begin{equation}
\label{ap1.6} \nu^{*}_{2}(\alpha)=\frac{1+\alpha}{d-1} \left[
\frac{d-1}{2}+\frac{3(d+8)(1-\alpha)}{16}+
\frac{4+5d-3(4-d)\alpha}{1024}c^{*}(\alpha)\right],
\end{equation}
\begin{equation}
\label{ap1.7}
c^{*}=\frac{32(1-\alpha)(1-2\alpha^{2})}{9+24d+(8d-41)\alpha
+30\alpha^{2}(1-\alpha)}\, .
\end{equation}
It is easily verified that $\kappa^{*}$ and $\eta^{*}$ tend to unity
when $\alpha$ goes to one, while $\mu^{*}$ and $\zeta^{*}$ vanish in
this limit.

\end{document}